\documentclass[review, onecolumn,english,american,prl,manuscript,superscriptaddress]{revtex4}
\usepackage[LGR,T1]{fontenc}
\usepackage[latin9]{inputenc}
\usepackage{color}
\usepackage{multirow}
\usepackage{makecell}
\setcitestyle{super}
\usepackage{amsmath}
\usepackage{amssymb}
\usepackage{graphicx}
\usepackage{textcomp}

\usepackage{array}
\newcolumntype{L}{>{\centering\arraybackslash}m{2cm}}

\makeatletter

\ProvideTextCommand{\~}{LGR}[1]{\char126#1}

\@ifundefined{textcolor}{}
{%
 \definecolor{BLACK}{gray}{0}
 \definecolor{WHITE}{gray}{1}
 \definecolor{RED}{rgb}{1,0,0}
 \definecolor{GREEN}{rgb}{0,1,0}
 \definecolor{BLUE}{rgb}{0,0,1}
 \definecolor{CYAN}{cmyk}{1,0,0,0}
 \definecolor{MAGENTA}{cmyk}{0,1,0,0}
 \definecolor{YELLOW}{cmyk}{0,0,1,0}
}

\usepackage[english]{babel}

\def\url#1{}
\addto\captionsenglish{%
}

\usepackage[none]{hyphenat}
\usepackage{multirow}

\usepackage{babel}

\makeatother

\usepackage{babel}

\begin{document}

\title{Integrating 2D Magnets for Quantum Devices: from Materials and Characterization to Future Technology}

\author{Han Zhong}
\author{Douglas Z. Plummer}
\affiliation{Department of Electrical and Computer Engineering, University of Florida, Gainesville, Florida 32611, USA.}
\author{Pengcheng Lu}
\affiliation{Department of Materials Science and Engineering, University of Florida, Gainesville, Florida 32611, USA.}
\author{Yang Li}
\author{Polina A. Leger}
\affiliation{Department of Electrical and Computer Engineering, University of Florida, Gainesville, Florida 32611, USA.}
%author list can be discussed later in our group meeting
\author{Yingying Wu}
\affiliation{Department of Electrical and Computer Engineering, University of
Florida, Gainesville, Florida 32611,
USA.}
\thanks{Correspond to: yingyingwu@ufl.edu}

\begin{abstract}

The unveiling of 2D van der Waals magnetism in 2017 ignited a surge of interest in low-dimensional magnetism. With dimensions reduced, research has delved into facile electric control of 2D magnetism, high-quality heterostructure design, and new device functionality. These atomically thin magnetic materials have spawned a burgeoning field known as 2D spintronics, holding immense promise for future quantum technologies. In this review, we comprehensively survey the current advancements in 2D magnet-based quantum devices, accentuating their role in manifesting exotic properties and enabling novel functionalities. Topological states, spin torques, voltage control of magnetic anisotropy, strain engineering, twistronics and designer interface will be discussed. Furthermore, we offer an outlook to guide their development in future CMOS and quantum hardware paradigms.
\end{abstract}
\maketitle

\newpage

\section{Introduction} % Polina in charge 
% 1. Brief overview of the significance of 2D materials in quantum technology; 2. Introduction to the concept of 2D magnets and their potential applications; 3. Purpose and scope of the review
% Rough draft introduction:

Two-dimensional (2D) magnets represent a groundbreaking class of materials that leverage the unique properties of 2D structures, such as flexibility and stackability, to advance the field of magnetism and spintronics. The discovery of ferromagnetism in 2D materials in 2017 marked a pivotal moment\cite{huang2017layer, gong2017discovery}. This breakthrough demonstrated that 2D intrinsic magnetic anisotropy and long-range magnetic order are achievable in certain van der Waals (vdW) materials, unlocking new potentials for technological applications. One of the key advantages of 2D magnets is the ability to manipulate layer thickness, which offers layer-dependent magnetic properties. This tunability, combined with interface and heterostructure engineering, allows for fine-tuning magnetic properties to meet specific needs \cite{recentprogress2dmagnets}. The reduced dimensionality of 2D magnets also facilitates the discovery of new phenomena that can enhance quantum effects and functionality. For example, the sharp 2D interface enables exploration of nanoscale quantum objects like 2D magnetic skyrmions\cite{wu2020neel,wu2022van,yang2020creation}. Thanks to the easy electric control of 2D ferroelectrics, tuning of skyrmion size down to a few nanometers was proposed in a In$_2$Se$_3$/Fe$_3$GeTe$_2$ 2D vdW heterostructure\cite{huang2022ferroelectric}. Transitioning from conventional memory technologies (e.g., SRAM, DRAM) to 2D multiferroic devices could yield comparable performance while introducing non-volatility, thereby reducing data retrieval time and power consumption \cite{wu2024nonreciprocal, zhang20242d}.

The ongoing discovery of new 2D magnets, facilitated by predictive models such as density functional theory, is broadening the scope of materials and their potential applications. These materials offer prospects for non-volatile memory, spin-based logic devices, and spin-dependent electronics.  Investigations in spintronics, magnonics, and microelectronics seek to capitalize on novel insights into exchange interactions and magnetic order. For example, antiferromagnetic materials can be manipulated by electric currents to attain read/write capabilities akin to ferromagnetic materials, while maintaining ultrafast speeds\cite{vzelezny2018spin}. Furthermore, the coexistence of magnetism and superconductivity in twisted 2D materials\cite{park2021tunable, gong2017discovery} hints at a profound interconnection between these phenomena, offering valuable insights for future quantum technologies\cite{recentprogress2dmagnets}.

This review paper delves into recent advancements in 2D magnetism, exploring its novel applications in non-volatile memories, magnetic sensors, and quantum technology. A comprehensive grasp of the properties and fabrication techniques of 2D magnets is indispensable for comprehending these applications. The paper endeavors to bridge solid-state principles from 3D to 2D within the realm of spintronics. Topics of discussion will encompass 2D topological orders in both real and momentum spaces, spin torques such as spin orbit torque (SOT) and spin transfer torque (STT), magnetoelectric coupling via voltage-controlled magnetic anisotropy, strain engineering with and without external pressure, twistronics applied to twisted 2D magnets for inducing exotic magnetic orders, and the creation of designer heterostructures to attain diverse high-quality interfaces. Beyond elucidating fundamental properties and effects, the review will explore practical applications in binary magnetic memory, biomedical devices, and quantum hardware. By offering a comprehensive overview of current research in 2D spintronics, the aim is to unveil potential pathways for technological innovation and provide insights into future hardware advancements.
\begin{figure}
\centerline{\includegraphics[width=1\textwidth]{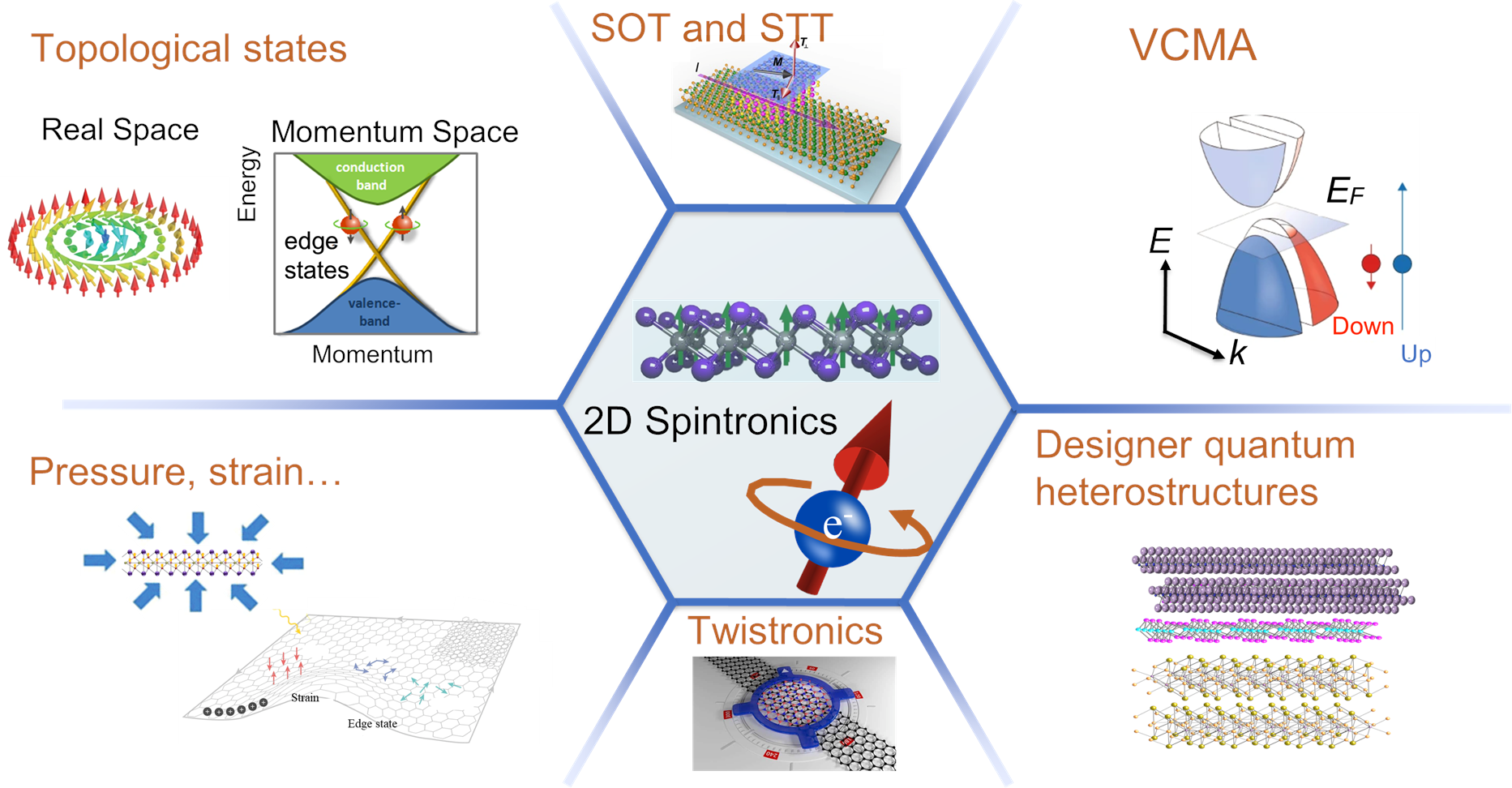}}
\vspace{-20pt}
\caption{Emerging 2D spintronics, driven by 2D magnets, sparking interest in various topics such as 2D topological orders, spin-torques, magnetoelectric coupling, strain engineering, twistronics, and designer heterostructures. }
\label{fig-intro}
\end{figure}

\section{Fundamentals of 2D Magnets} \label{materials}%Han Zhong in charge
% 1.{ Definition and characteristics of 2D magnets; 2. Comparison between bulk magnets and 2D magnets; 3. Overview of different types of 2D magnets; 4.Synthesis techniques for 2D magnets}
%Please help add references
2D magnets can be atomically thin, sometimes just a single layer of atoms $\sim$ 0.8 nm, while preserving the magnetism. They exhibit remarkable magnetic anisotropy, leading to quantum phenomena not observed in their bulk counterparts. What sets them apart from traditional bulk magnetic materials are their unique features. These include quantum confinement, which arises from carriers being confined to a 2D plane, as well as mechanical flexibility due to their ultrathin thickness. Additionally, they possess enhanced electrical, thermal and optical sensitivity\cite{ma2021tunable,zhang2021two,chen2023effect}, rendering them highly responsive to external stimuli such as electric fields, laser control, chemical alterations, and mechanical strain. Moreover, their electronic and magnetic properties can be finely tuned through external adjustments like electric gating\cite{tang2023continuous,wu2023gate} and strain\cite{ren2023strain,cenker2022reversible}. These distinctive characteristics open up new avenues for 2D magnets in ultracompact devices. For instance, their 2D properties enable applications in quantum computing\cite{kezilebieke2020topological,kurebayashi2022magnetism} and the development of highly sensitive magnetic sensors\cite{jimenez2020magnetic,khan2020recent}.
%pay attention to the logics, make it logically smooth: 2D -> quantum confinement -> difference from bulk, like high mobility, easy electric control... -> different groups of 2D magnets, ferromagnet, antiferromagnet, multiferroic, quantum spin liquid crystals... -> how mechanical exfoliation used to obtain thin films, what is the lateral size, different approaches, like mechanical exfoliation with scotch tape, gold-assisted exfoliation, Al2O3- coated exfoliation, MBE growth, sputtering? ...

A diverse array of 2D magnets has been uncovered (Table \ref{table-material}), spanning a wide spectrum of materials, from metals to semiconductors to insulators, each exhibiting a range of magnetic behaviors including ferromagnetic, antiferromagnetic, and multiferroic properties. Beginning with binary transition metal halides like CrI$_3$ and CrBr$_3$, structured as layered vdW solids, they are renowned for their either ferromagnetic or antiferromagnetic interlayer order depending on layer stacking. In contrast, MnBi$_2$Te$_4$ showcases interlayer antiferromagnetic order alongside topological properties, extensively studied for phenomena such as the quantum anomalous Hall effect \cite{deng2020quantum}, which manifests quantized Hall resistivity at elevated temperatures up to 45 K \cite{bai2024quantized,liu2021magnetic,ge2020high} under applied magnetic fields. Ternary transition metal compounds such as Cr$_2$X$_2$Te$_6$ (X = Si, Ge, Sn) and MPX$_3$ (M = transition metal; X = S, Se, Te) represent materials with intricate interactions between constituents, resulting in diverse magnetic and electronic properties. For example, Cr$_2$Ge$_2$Te$_6$ is reported as an insulator with a band gap of $\sim$0.7 eV in bulk\cite{ji2013ferromagnetic} and order in a
Heisenberg-type ferromagnetism below T$_\textrm{C}$ 61 K. The two-probe resistance of this material at room temperature usually in the order of $10^5\, \Omega$ to $10^6\,\Omega$. This electrically resistive nature poses challenges to transport characterization. Furthermore, emerging 2D rare-earth and f-electron magnets show promise for high-temperature and high-density magnetic storage, like DyOCl\cite{tian2021dyocl} as an A-type rare-earth antiferromagnet and gadolinium halide GdX$_2$ (X = I, Br, Cl, F)as high-temperature 4f-electron ferromagnets\cite{you2022gadolinium}. 
\begin{table}[!ht]
\small
 \begin{center}
 \begin{tabular}{|c|c|c|c|c|}
  \hline
  Materials& \makecell{Bandgap\\(eV)} & \makecell{Curie/N\'eel Tempe\\(K)}& Magnetism & \makecell{Stability\\(air)} \\ \hline
 VSe$_2$ & .055\cite{bonilla2018strong} & 400-600 & Ferro & Yes \\ \hline
  Fe$_3$GaTe$_2$  & 0 & 350-380\cite{zhang2022above} & Ferro & Yes \\ \hline
Fe$_5$GeTe$_2$ & 0 & 270-330\cite{yuan2023modulating} & Ferro & Yes\\ \hline
CrTe$_2$ & 0.37-0.50 & 310\cite{liu2022structural} & Ferro & Yes\\ \hline
%VSe$_2$ & 55 meV (150 K)\cite & 292\cite{fuh2016newtype} & Ferro & No \\ \hline
Fe$_3$GeTe$_2$ & 0 & 220-230\cite{deng2018gate,wu2020large} & Ferro & Yes\\ \hline
Ta$_3$FeS$_6$ & 0 & 80\cite{su2020air} & Ferro & Yes \\ \hline
Cr$_2$Ge$_2$Te$_6$ & 0.7\cite{ji2013ferromagnetic} & 61 & Ferro & Yes \\ \hline
% VI$_3$ & 0.6 & 50 & Ferro & No \\ \hline
CrBr$_3$ & 1.68-2.1\cite{wu2022optical} & 30\cite{singh2022room} & Ferro & No \\ \hline
% not sure whether this exist Cr$_2$Si$_2$Te & 0.2 & 33 & Ferro & Yes \\ \hline
CrI$_3$ & 1.3-1.4\cite{su2021enhancing} & 14-45 & Ferro & No \\ \hline
%add reference for FeCl2, VI3, CrBr3... you go to Google Scholar find the paper, click "cite on the bottom, click bibtex on the bottom. It will automatically generate bibtex
% FeCl$_2$ & 4.1 & 17 & Ferro & No \\ \hline
%CrBiSbTe & NA (insulator) & 13 & Ferro & Yes \\ \hline
GdI$_2$ & 1 & 251\cite{liu2021two} & Ferro & Yes \\ \hline
CrSBr & 1.42-1.47\cite{bo2023calculated} & 132 & Antiferro & Yes \\ \hline
FePS$_3$ & 1.77\cite{wu2022tunable} & 117 & Antiferro & Yes \\ \hline
MnPSe$_3$ & 2.32\cite{zhang2016mnps3} & 70 & Antiferro & Yes \\ \hline
CrPS$_4$ & 1.4\cite{susilo2020band} & 40 & Antiferro & Yes \\ \hline
MnBi$_2$Te$_4$ & 0.55\cite{trang2021crossover} & 25 & Antiferro & No \\ \hline
CrCl$_3$ & 2.6\cite{mastrippolito2021emerging} & 14 & Antiferro & No \\ \hline
DrOCl &5.72 & 10 & Antiferro & Yes \\ \hline
CuCrP$_2$S$_6$ & 1.2\cite{guo2024electronic} & 145\cite{wang2023electrical} & Multiferroic &  Yes \\ \hline
NiI$_2$ & 1.11-1.23\cite{liu2020vapor} &21\cite{song2022evidence} & Multiferroic & No \\ \hline
\end{tabular}
 \end{center}
 \vspace{-20pt}
 \caption{A list of 2D magnets and their properties.}
 \label{table-material}
\end{table}

Synthesis techniques wield significant influence over the quality and properties of 2D magnets, as listed in Table \ref{table-grow}. Mechanical exfoliation, a commonly employed method, typically yields the highest quality but faces limitations in scalability, with the lateral size in the range of micrometers to a hundred micrometers, with potential oxidation during the process posing risks to material quality. New exfoliation methods, such as universal gold-assisted exfoliation, extend fabrication capabilities by enabling the production of monolayers on a millimeter to centimeter scale\cite{huang2020universal}. This method has proven to be successful for over forty varieties of vdW materials, including Fe$_3$GeTe$_2$.The exfoliated monolayer flakes can be intactly transferred onto arbitrary substrates after removing gold layer by KI/I$_2$ etchant. The removal of gold layer may bring unexpected contamination in samples prepared for electrical and optical measurements and thus reduce their performances when removing gold films with chemical solvents in additional steps. The alternative way is to keep gold layer thickness below 2-3 nm, in the electrically insulating region. Additionally, Al$_2$O$_3$-coated exfoliation, achieved through atomic layer deposition (ALD), provides enhanced thickness and uniformity control while safeguarding material integrity \cite{deng2018gate}. Removal of the Al$_2$O$_3$ layer in solvent may cause contamination to the 2D surface. This layer can be retained and serve as a dielectric layer for electric gating. Liquid phase exfoliation (LPE) stands out for its simplicity and controllable sizing, featuring efficient and stable three-stage sonication-assisted exfoliation for mass production of high-integrity few- and single-layer Fe$_3$GeTe$_2$ nanoflakes, with lateral sizes reaching up to 103 $\mu$m \cite{ma20222d}. This method offers pathways for diverse applications, including printed electronics and composite materials. Vapor deposition technique has been utilized for the growth of NiI$_2$ on substrates such as SiO$_2$/Si and h-BN down to monolayer thickness, enabling the production of high-purity solid materials through vapor-phase precursor reactions on heated substrates, prized for its ability to generate uniform, high-quality coatings \cite{liu2020vapor}. Molecular beam epitaxy (MBE), a meticulously controlled vacuum deposition technique, facilitates atomic layer-by-layer deposition of crystalline materials with low defect densities. MBE-grown Fe$_3$GeTe$_2$ on Bi$_2$Te$_3$ substrates has demonstrated room-temperature ferromagnetism, showcasing its potential for applications in quantum devices. Later a high SOT efficiency at room temperature was reported in this heterostructure, with details shown in Section \ref{sec:SOT}. Sputtering, which deposits thin films onto substrates using high-energy ion particles, is known for its uniformity across various materials and substrate shapes. While extensively used for 2D non-magnetic materials like MoS$_2$ \cite{zhang2019van}, its potential applications to 2D magnets are yet to be fully explored.

\begin{table}[!ht]
\small
 \begin{center}
 \begin{tabular}{|c|c|c|c|c|}
  \hline
 Method & Size & Thickness & Materials & Challenge\\ \hline
 mechanical exfoliation & $\mu$m & varied & universal & small scale monolayers\\ \hline
 gold-assisted exfoliation & mm-cm & monolayer & universal & solvent to remove gold layer \& contaminiation \\ \hline
 Al$_2$O$_3$-coated exfoliation & $\mu$m & monolayer & Fe$_3$GeTe$_2$ & Al$_2$O$_3$ attached \\ \hline
 LPE & $\mu$m & controllable & Fe$_3$GeTe$_2$ & potential doping  \\ \hline
 vapor deposition & $\mu$m & few-layer & NiI$_2$ & uniform monolayers \\ \hline
 MBE & cm & few-layer & Fe$_3$GeTe$_2$ & high-quality monolayer \\ \hline
\end{tabular}
 \end{center}
 \vspace{-20pt}
 \caption{Comparison of methods using exfoliation or in-situ growth of 2D magnets.}
 \label{table-grow}
\end{table}

%\begin{figure*}[ht]
%\begin{centering}
 %\includegraphics[width=0.85\textwidth]{fig1.jpg} 
%\par\end{centering}
%\caption{Illus. }\label{Fig1} 
%\end{figure*}

\section{Magnetic properties of 2D Materials}\label{property} % Pengcheng in charge
%1.Magnetic ordering and spin configurations in 2D materials; 2. Magnetic anisotropy and its role in 2D magnets; 3. Factors influencing magnetic properties (e.g., defects, strain); 4. Magnon.

Due to their atomic thinness, 2D materials provide an opportunity to address concerns regarding the dilution of magnetic interactions observed in bulk materials, offering stronger and more precise magnetic responses across various domains. The Mermin-Wagner limit \cite{mermin1966absence} dictates that long-range magnetic order is unattainable at finite temperatures in an ideal 2D Heisenberg model with continuous symmetry and short-range interactions. Magnetic anisotropy provides a solution. Referring to the directional dependence of a material's magnetic properties, magnetic anisotropy can stabilize magnetic order in a low-dimensional system by breaking the continuous rotational symmetry. Anisotropic properties can help create a magnon excitation gap \cite{gong2019two}, thereby counteracting thermal fluctuations and facilitating long-range order. Ideally, this results in a finite Curie temperature T$_\textrm{C}$, ensuring the existence of 2D magnetism.

\begin{figure}
\centerline{\includegraphics[width=0.6\textwidth]{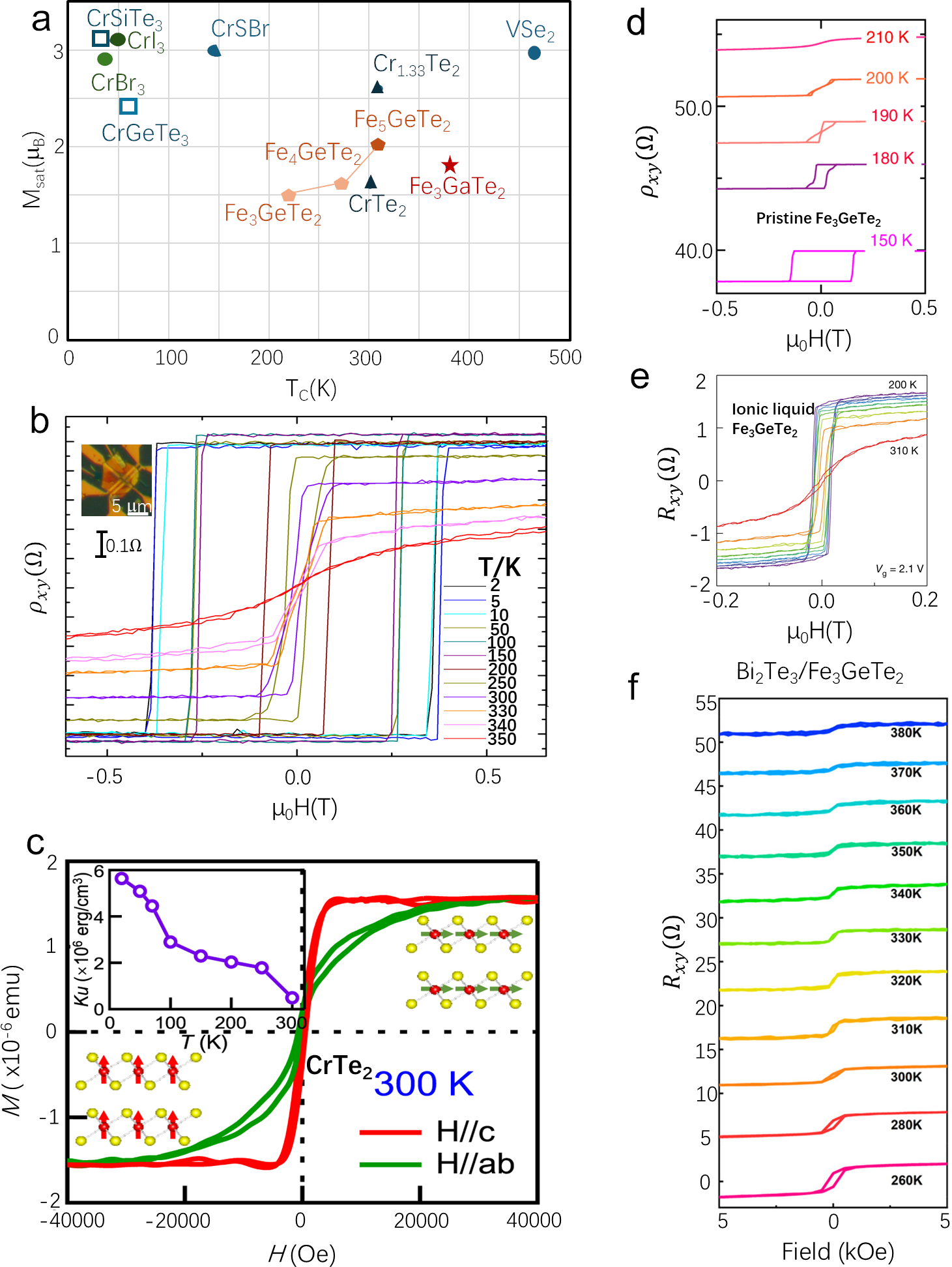}}
\caption{Reported room-temperature 2D magnets. (a) Current extensively studied 2D magnet materials, with Fe$_3$GaTe$_2$ showing the highest Curie temperature. (b) Anomalous Hall resistance characterization of few-layer Fe$_3$GaTe$_2$, showing Curie temperature above 340 K. (c) Room-temperature ferromagnetism in CrTe$_2$, revealed by SQUID measurement\cite{zhang2021room}. (d) Pristine few-layer Fe$_3$GeTe$_2$ showing a Curie temperature around 200 K\cite{wu2020neel}. (e) Ionic liquid gating adopted to tune the carrier density in Fe$_3$GeTe$_2$\cite{deng2018gate}, leading to room-temperature magnetism. (f) Epitaxy growth of few-layer Fe$_3$GeTe$_2$ on Bi$_2$Te$_3$ substrates indicating an above-room-temperature ferromagnetism\cite{wang2020above}.  }
\label{fig-RT}
\end{figure}

The presence of strong magnetic anisotropy can facilitate the development of room-temperature 2D magnets, which are pivotal for advancing various technological applications. These magnets offer practical advantages, including enhanced energy efficiency, simplified device design, and compatibility with existing magnetic memory infrastructure. Furthermore, their stability at room temperature enables seamless integration into everyday electronics, eliminating the need for complex and costly cooling systems. This makes such materials economically viable for widespread adoption in consumer electronics, telecommunications, and beyond. Among the discovered room-temperature 2D magnets depicted in Fig. \ref{fig-RT}, VSe$_2$ is a high-temperature ferromagnet. By varying the substrate\cite{yu2019chemically,kezilebieke2020electronic} and strain\cite{sheng2021magnetic}, its T$_\textrm{C}$ can be varied between 400 K and 600 K. Intrinsic magnets like CrTe$_2$ and Fe$_3$GaTe$_2$ were also studied. Their magnetic properties are primarily characterized through anomalous Hall resistivity measurements (Figs. \ref{fig-RT}b \& d-f) or SQUID measurements (Fig. \ref{fig-RT}c) on the magnetization. Fe$_3$GaTe$2$, for instance, is reported to be a 2D vdW ferromagnet with the high T$_\textrm{C}$ reaching up to 380 K depending on Fe vacancy, a temperature suitable for ambient operation even considering heating in CMOS environments. Conversely, Fe$_3$GeTe$_2$ exhibits ferromagnetism below room temperature, while its Curie temperature can be adjusted through doping. Techniques such as ionic liquid gating can enhance the Curie temperature of few-layer Fe$_3$GeTe$_2$ to above room temperature, while MBE growth of this material onto Bi$_2$Te$_3$ can further augment its Curie temperature. 

\subsection{Electrical characterizations}
This process involves measuring various electrical parameters, such as resistivity, conductivity, and Hall effect, to gain insights into the material's magnetic behavior and electronic structure. Techniques like magneto-transport measurements can reveal how electrical resistance changes with an applied magnetic field, providing information about phenomena such as anisotropic magnetoresistance (AMR), tunneling magnetoresistance (TMR) and anomalous Hall resistance (AHR). 

%AMR was studied in both 2D ferromagnets and antiferromagnets\cite{fang2022quasi,liu2019anisotropic}, promising for polarization sensitive spintronics. TMR was reported as large as 10,000\% in the direction perpendicular to the CrI$_3$ crystalline planes\cite{wang2018very}. 
\textbf{A1: Curie temperature characterization}
Most materials-based work using transport measurement on AHR showing a hysteresis loop of ferromagnets, where the AHR is related to magnetization\cite{wu2020neel,wu2022manipulating,wu2022van}. This was well established and sometimes not applicable to antiferromagnet case.

\textbf{A2: Voltage control of magnetic anisotropy (VCMA)}
Additionally, electrical characterization helps in assessing the influence of external stimuli, such as electric voltage on the magnetic state of the material.
Multiferroic control of 2D ferromagnetism was realized using $\pm$5 V across the Cr$_2$Ge$_2$Te$_6$/polymer heterostructure, where can magnetic hysteresis loop can be opened and closed. The magnetic modulation is non-volatile, and is observed in bilayer, trilayer and four-layer Cr$_2$Ge$_2$Te$_6$\cite{liang2023small}.  

\textbf{A3: New transistors}
This also gives rise to new transistor types, like spin tunnel field effect transistors\cite{jiang2019spin}. The devices exhibit an ambipolar behaviour and tunnel conductance that is dependent on the magnetic order in the CrI$_3$ tunnel barrier. The gate voltage switches the tunnel barrier between interlayer antiferromagnetic and ferromagnetic states under a constant magnetic bias near the spin-flip transition, thus effectively and reversibly altering the device between a low and a high conductance state, with large hysteresis. Understanding these interactions is crucial for designing devices that exploit spin-orbit coupling or spin-transfer torque for efficient data storage and processing.

\textbf{A4: Topological properties}
As we have mentioned in Fig. \ref{fig-intro}, topological properties in both real and momentum space are allowed in 2D spintronics. As shown in Section \ref{materials}, MnBi$_2$Te$_2$ is a Chern insulator\cite{wang2020topological}, showing topology with momentum-space Berry phase, leading to quantization of AHR. Meanwhile, topology in real space shown as magnetic skyrmions was also reported in 2D magnets-based heterostructure\cite{wu2020neel,wu2022van,zhang20242d}, extensively studied in Fe$_3$GeTe$_2$ and Cr$_2$Ge$_2$Te$_2$ case. The sharp nature  of 2D interface can host two groups of magnetic skyrmions at one interface, which is absent from epitaxy-grown or sputtering films.  

\textbf{A5: Strain engineering}
Tensile strain is a proven method for tuning the fundamental properties of three-dimensional bulk quantum materials and 2D thin films. For vdW magnets, recent works have highlighted the tantalizing opportunities for the strain control of magnetic properties such as ordering temperature and coercive field from hysteresis loop in Fe$_3$GeTe$_2$\cite{wang2020strain}, with a dramatic increase of the coercive field by more than 150\% with an applied strain of 0.32\%. 

\subsection{Optical properties}
Since exfoliated 2D magnets often have a small lateral size around $\mu$m, optical detection with a fine resolution are required to probe their optical properties. MOKE measures the change in polarization of reflected light in response to magnetization, with a resolution of 1-2 $\mu$m. It can help reveal critical information about the materials magnetic domains\cite{pan2022efficient}, anisotropy and phase transitions\cite{wu2020neel}. Additionally, 2D magnets can exhibit unique excitonic effects and spin-dependent optical transitions, which are influenced by their reduced dimensionality and strong spin-orbit coupling. A reversible strain-induced antiferromagnetic-to-ferromagnetic phase transition and strain control of the out-of-plane spin-canting process were read through polar reflective magnetic circular dichroism (RMCD) in CrBrS at zero magnetic field\cite{cenker2022reversible}. 
These characteristics enable the development of novel optoelectronic devices, such as spin-LEDs and magnetic photodetectors, which can operate with high efficiency and sensitivity. Furthermore, the ability to manipulate magnetic properties using light, known as all-optical switching\cite{zhang2022all,dabrowski2022all}, opens new possibilities for ultrafast data storage and processing technologies. 

\subsection{Magnetic imaging}
\begin{figure}
\centerline{\includegraphics[width=1\textwidth]{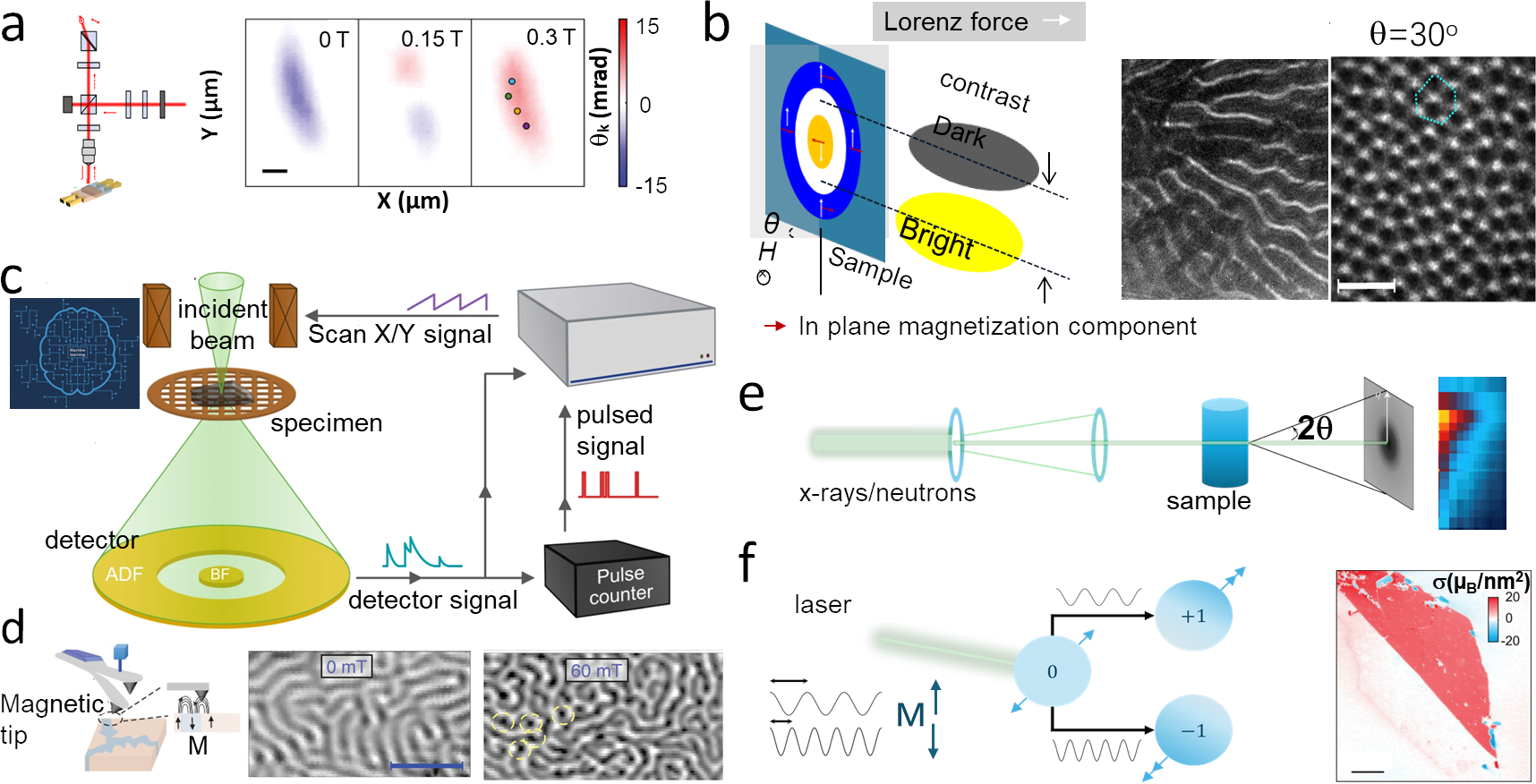}}
\vspace{-20pt}
\caption{Magnetic imaging methods. (a) MOKE setup\cite{wu2020large}, with an examples showing mapping of magnetization on CrI$_3$ layers\cite{huang2017layer}. (b) L TEM exhibiting contrast of dark and bright, examples including magnetic domain and magnetic skrymions imaging\cite{wu2020neel}. (c) 4D STEM integrated with machine learning offering high resolution\cite{peters2023electron,kalinin2022machine}. (d) MFM using a magnetic tip, with examples showing magnetic domains of Cr$_2$Ge$_2$Te$_6$ and Fe$_3$GeTe$_2$\cite{wu2022van}. (e) SANS mapping of magnetic textures\cite{jeffries2021small}, with an example of CrSBr\cite{rybakov2024probing}. (f) NV center mapping of magnetization, showing an example on CrI$_3$ magnetization\cite{thiel2019probing}. } 
\label{fig-image}
\end{figure}

Magnetic imaging encompasses various techniques to visualize and understand magnetic structures at different scales and resolutions (Fig. \ref{fig-image}a \& Table \ref{table-imaging}). Among most techniques, MOKE (Fig. \ref{fig-image}a) was extensively adopted for magnetic domain imaing, with a low resolution of 1-2 $\mu$m. Lorentz transmission electron microscopy (L TEM) exploits electron phase shifts caused by magnetization within samples (Fig. \ref{fig-image}b), allowing the observation of magnetic domain structures with high spatial resolution. It is regarded as the most efficient method because of its ultrahigh magnetic spatial resolution (down to 2 nm)\cite{yu2012skyrmion,zhao2016direct}. The fine magnetic structure of a skyrmion was first observed in real-space using L TEM\cite{yu2011near}. It was also used in the first observation of 2D skyrmions in vdW WTe$_2$/Fe$_3$GeTe$_2$ heterostructure\cite{wu2020neel}.  
4D scanning transmission electron microscopy (STEM), a variant of TEM, provides detailed atomic-level imaging by scanning a focused electron beam over the sample (Fig. \ref{fig-image}c), useful in correlating magnetic properties with atomic structure. Semi-quantitative maps of the magnetic structures can be obtained with a spatial resolution nearly equal to the size of the electron probe or 10-20 nm. Aberration corrected STEM can have a spatial resolution lower than 1 nm\cite{nago2016evolution}. It was proposed machine learning can be integrated with 4D STEM to  automate experiments and novel scanning modes\cite{kalinin2022machine}. Fully realizing this potential necessitates significant developments on multiple levels, from instrumental platforms to common workflows, shared data and codes.  
In magnetic force microscopy (MFM), the magnetic stray field from nanoscale features is detected by a scanning nanoscale magnetic tip (Fig. \ref{fig-image}d). This technique has been applied to image magnetic domain and skyrmions in Fe$_3$GeTe$_2$ and Cr$_2$Ge$_2$Te$_6$ thin films\cite{wu2022van}. 
Small-Angle Neutron Scattering (SANS) probes magnetic correlations in materials by scattering neutrons at small angles, suitable for studying bulk magnetic structures and dynamics on the nanoscale (Fig. \ref{fig-image}e). It was used to show the formation of short-range magnetic regions in CrSBr with correlation lengths at the antiferromagnetic ordering temperature (T$_N$ $\sim$ 140 K)\cite{rybakov2024probing}.
Nitrogen-Vacancy (NV) centers (Fig. \ref{fig-image}f) in diamond offer a quantum-based approach to magnetic imaging, enabling high-sensitivity detection of magnetic fields at the nanoscale through the interaction of NV electron spins with external magnetic fields. This technique has been applied to image magnetic domains in CrI$_3$, showing a spatial resolution of $\sim$50 nm\cite{thiel2019probing}. Other techniques like scanning transmission
X-ray microscopy (STXM) and X-ray magnetic circular dichroism (XMCD) are adopted for imaging in Fe$_3$GeTe$_2$ and Fe$_5$GeTe$_2$ system\cite{park2021neel,schulz2023direct}. 

\begin{table}[!ht]
\small
 \begin{center}
 \begin{tabular}{|c|c|c|c|c|}\hline
 Method & spatial resolution & Pros & Best used for &  vdW magnets studied \\ \hline
 MOKE & 1-2 $\mu$m& easy access & large-scale domain & CrI$_3$\cite{dabrowski2022all}\\ \hline
 L TEM & 2 nm & \makecell{High resolution\\easy acces} & thin film & \makecell{Fe$_3$GeTe$_2$\cite{wu2020neel,ding2019observation,wang2020characteristics}\\Cr$_2$Ge$_2$Te$_6$\cite{han2019topological,mccray2023direct} }\\ \hline
4D STEM & 1 nm & \makecell{High resolution\\no post image process} & thin film & Co-doped Fe$_5$GeTe$_2$\cite{shao2022room} \\ \hline
MFM & 5-10 nm & Electrical signal allowed & thin film & \makecell{Cr$_2$Ge$_2$Te$_6$\cite{wu2022van,guo2021multiple}\\Fe$_2$GeTe$_2$\cite{leon2016magnetic}} \\ \hline
SANS & 0.5-1 nm & High resolution & bulk or film  & DyTe$_3$\cite{akatsuka2024non} \\ \hline
NV center & 50 nm & Sentitive& Ultrathin film &\makecell{VI$_3$\cite{broadway2020imaging} \\CrI$_3$\cite{laraoui2022opportunities,thiel2019probing}}\\ \hline
\end{tabular}
 \end{center}
 \vspace{-20pt}
 \caption{Comparison of magnetic imaging techniques.}
 \label{table-imaging}
\end{table}

\subsection{Magnon-phonon interaction}
The interaction between magnons and phonons in 2D magnets is a fascinating area of study that has significant implications for both fundamental physics and practical applications\cite{chen2021electrically,qi2023giant}. Magnons, which are quasiparticles representing collective excitations of electron spins, and phonons, which are quasiparticles representing lattice vibrations, can strongly influence each other's behavior in 2D materials due to their reduced dimensionality and enhanced surface effects. This magnon-phonon coupling can affect the thermal\cite{chen2024magnon,yang2023spin} and magnetic properties\cite{delugas2023magnon} of 2D magnets, leading to phenomena such as spin Seebeck effect, where a temperature gradient generates a spin current. Understanding these interactions is crucial for developing efficient spintronic devices, as they can impact the propagation of spin waves and the dissipation of energy. Moreover, manipulating magnon-phonon interactions could enable new ways to control magnetic states and develop low-power, high-speed information processing technologies. Advances in this field could thus provide novel insights into the dynamic processes within 2D magnets and drive innovations in areas ranging from quantum computing to thermoelectrics.

\section{Integration of 2D Magnets in Quantum Devices}
\subsection{Spintronics and magnetic memory devices}
% the overall paragraph looks good. Then you need to add more detailed info on 2D spintronic device. You can add a table listinig 2D SOT efficiency compared to other thin-film SOT device; a table listing pros and cons of 2D and other STT device; then illustrate how VCMA can be faciliatated in the 2D case, by showing figures and descriptions. 
%SOT, STT, spin-based logic device, spin FETs 
Spintronics, arising at the intersection of magnetism and electronics, has experienced significant advancements through the integration of 2D magnets. Spin-orbit torque (SOT) and spin-transfer torque (STT) mechanisms have played pivotal roles in enhancing the functionality and efficiency of spin-based logic devices, such as spin field-effect transistors (FETs). By harnessing 2D magnets in these devices, improved controllability, capacity, and reduced energy consumption have been achieved, revolutionizing the capabilities of modern data storage and processing. This section endeavors to elucidate the 2D magnetic properties that underpin this transformative progress in magnetic memory, offering insights into their potential for future applications.

\subsubsection{Spin-Orbit Torque (SOT)}\label{sec:SOT}
\begin{figure}
\centerline{\includegraphics[width=0.6\textwidth]{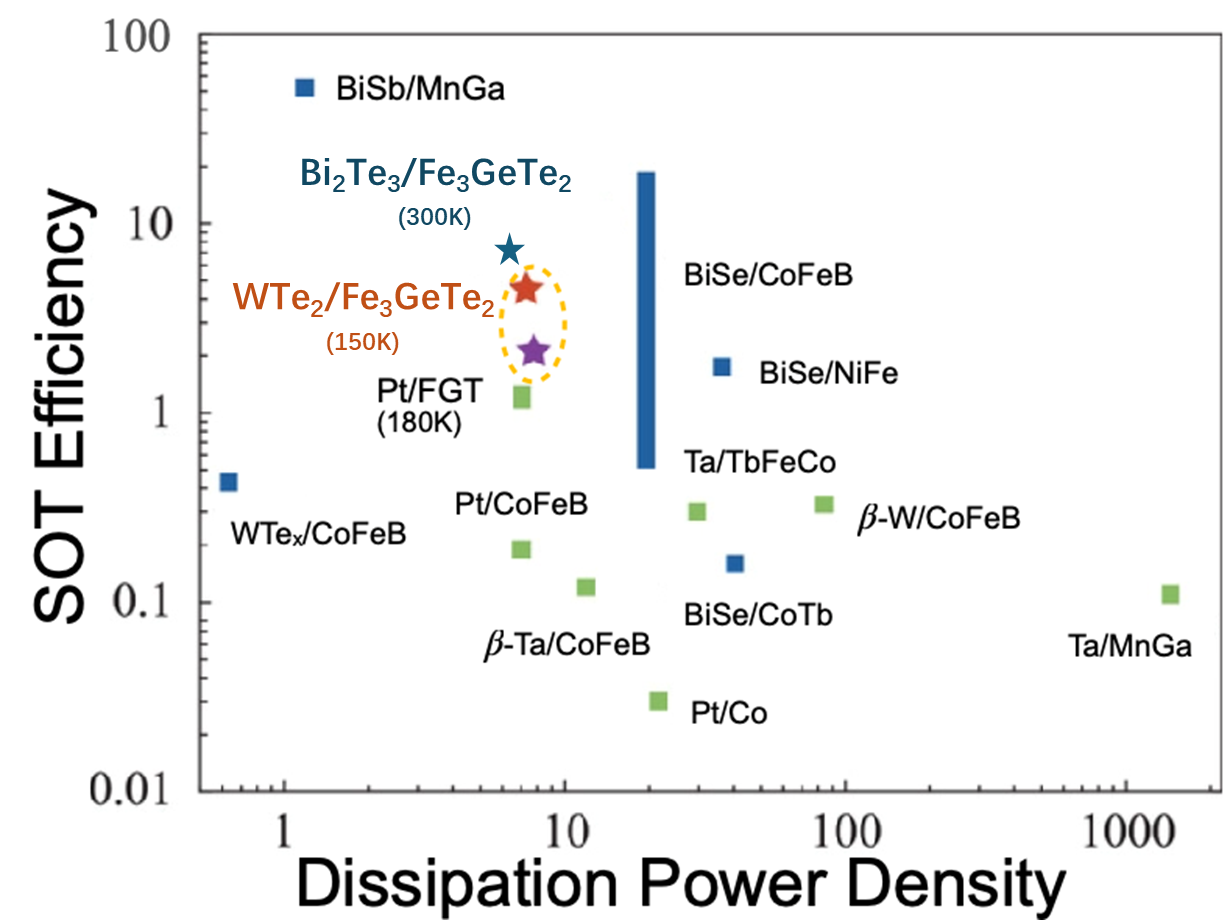}}
\vspace{-20pt}
\caption{SOT in 2D magnet-based heterostructure showing a high efficiency with relatively low dissipation power density, compared to other system.}
\label{fig-SOT}
\end{figure}

In conventional information systems, signals are switched between binary states 0 and 1, generated by different magnetic fields. Such an approach relies on magnets or current coils that occupy a considerable amount of space and energy. Under the trend of minimizing device fabrication, SOT, serves as a fast switch based on magnetic structures. The SOT structure consists of a non-magnetic conductor and a ferromagnetic heterostructure (Fig. \ref{fig-sensor}a). In such a system, the spin current generated by the spin Hall effect (SHE) will pass through the adjacent ferromagnetic layer when an electrical current is applied, inducing SOT. This technology allows the device to become non-volatile, allowing it to store the data when voltage is off. SOT can receive angular momentum through four different ways, i.e. magnetic order, carrier spins, orbital moments, and atomic lattice. SOT usually can be categrorized into Spin Hall Torque and Rashba-Edelstein Torque. The magnetization due to spin Hall torque runs toward the interface plane, and Rashba-Edelstein Torque rotates about the interface plane. Fig. \ref{fig-SOT} listed reported SOT efficiency in some structures. Compared to non-vdW system, SOT in vdW heterostructures are in high efficiency and low dissipation power region. Room-temperature SOT driven magnetization switching in an all-vdW heterostructure was realized using an optimized epitaxial growth. The topological insulator Bi$_2$Te$_3$ not only raises the Curie temperature of Fe$_3$GeTe$_2$ through interfacial exchange coupling but also works as a spin current source allowing the Fe$_3$GeTe$_2$ to switch at a low current density of $\sim$2.2$\times$10$^6$ A/cm$^2$. The SOT efficiency is $\sim$2.69, measured at room temperature\cite{wang2023room}. In the case of WTe$_2$/Fe$_3$GeTe$_2$, a high SOT efficiency of 4.6 and a switching current density of 3.90 $\times$ 10$^6$ A/cm$^2$ at 150 K was reported\cite{shin2022spin}. 

\subsubsection{Spin-Transfer Torque (STT)}

\begin{figure}
\centerline{\includegraphics[width=0.7\textwidth]{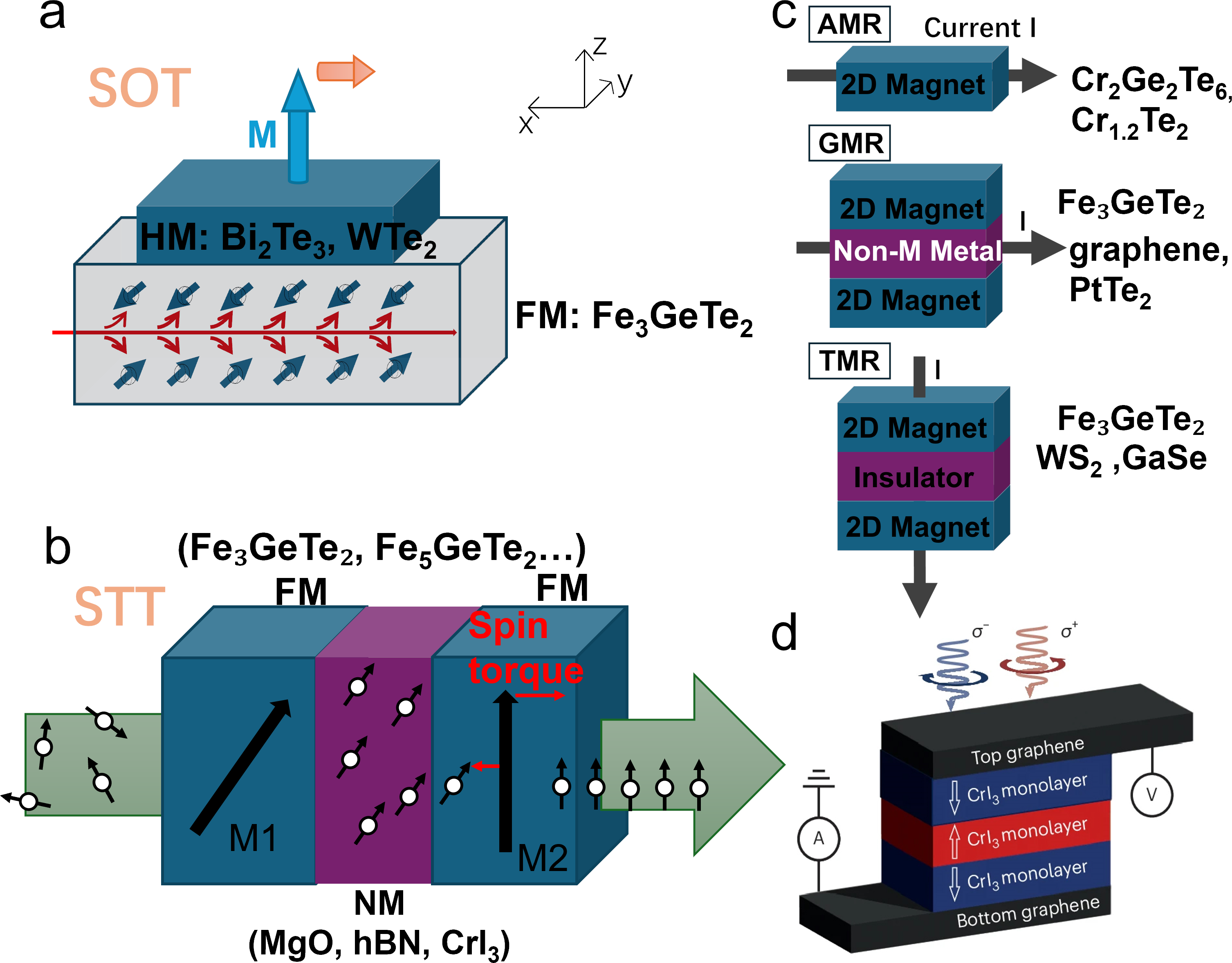}}
\vspace{-20pt}
\caption{Illustration of SOT, STT, magnetism sensor mechanism and example. (a) 2D SOT made with vdW materials. (b) STT using vdw materials for ferromagnetic layer and non-magnetic spacer. (c) Traditional magnetic sensor mechanism. (d) Sensitive photodector on CrI$_3$\cite{gish2024van}.}
\label{fig-sensor}
\end{figure}

A STT device consists of two ferromagnetic layers separated by a spacer\cite{dey2021spin} (Fig. \ref{fig-sensor}b). The first ferromagnetic layer has a fixed magnetization direction and the second ferromagnetic layer has a free magnetization that can be adjusted by the spin-polarized current. The non-magnetic layer serves as a spacer, allows the flow of spin-polarized electrons without altering their spin direction.  When the electrons enter the device, they first pass through the fixed ferromagnetic layer and become spin-polarized. This means that most electrons will have their spins aligned along the direction of the first ferromagnetic layer. Later, these spin-polarized electrons enter the free ferromagnetic layer, and they exert a torque on its magnetization\cite{dey2021spin}. STT device has huge potential in advanced memory technologies, particularly magnetic random-access memory (MRAM).
The direction of these magnetic domains represents binary data (0 and 1) and remains stable when there are no external power supplies, called non-volatility. The key component of STT-MRAM is the magnetic tunnel junction (MTJ), which consists of an insulating layer sandwiched by two ferromagnetic layers. After the current is applied, the spin-polarized current will pass through the MTJ and exert torque on the magnetic moments in the free layer. This will force them to align with the direction of the current. Adoption of 2D magnets in STT allows ultracompact devices, considering their uniform thickness down to $\sim$ 0.8 nm. Examples\cite{zhang2021recent} have been shown on using vdW magnets as spacer\cite{yan2020significant,wang2018very,song2019voltage} (like CrI$_3$) or ferromagnetic metal\cite{li2019spin,li2021large,zhao2023magnetoresistance} (like Fe$_5$GeTe$_2$). 

\subsection{Magnetic sensors}
%need specific examples, you can cite other papers. need to illustrate how it works in the sensor application, what is the advantage, what is the motivation people combine 2D [DOUG]
The integration of 2D magnets has profoundly influenced the landscape of magnetic sensing technologies; enabling the realization of highly sensitive and compact sensors including AMR, GMR, and TMR which have been leveraged to enhance the performance of magnetic sensors(Fig. \ref{fig-sensor}c), facilitating diverse applications ranging from automotive systems to biomedical diagnostics. Additionally, the emergence of vdW room-temperature ferromagnetic materials %including VSe$_2$ \cite{bonilla2018strong}, 
has spurred related advancements in nano-sensor technology. For example, the vdW magnetic semiconductor CrI$_3$ in a tunnelling geometry has been used to create a photodetector that is sensitive to the helicity of circularly polarized light\cite{gish2024van}. This offers unprecedented opportunities for miniaturization and integration with emerging platforms.

\textbf{Anisotropic Magnetoresistance}
AMR is a phenomenon where the resistance of a specific material is influenced by the angle between an applied current and the direction of magnetization in the material being used. It is important to note that the type of ferromagnetic material also plays a key role in the effect, with metals like nickel experiencing over three times the magnetoresistance when compared to iron. This is because the effect is a byproduct of scattering that occurs in conducting electrons with uncompensated spins. It is also be paramount to recognize that the resistance of these simple ferromagnets varies in a nonlinear dependence on the applied field \cite{popovic1996future}. This is compounded with the fact that when the applied field (separate from the total magnetic anisotropy field) is small, the sensitivity of the sensor suffers, with the sensor much more likely to produce false negatives or misidentify what it is observing\cite{khan2021magnetic}. In the case of Cr$_2$GeTe$_6$, the field-dependent resistance shows a positive or negative magnetoresistance effect when magnetic field H is switched to different directions\cite{liu2019anisotropic}. Cr$_{1.2}$Te$_2$ flake reveals a room-temperature remarkable 4-fold AMR effect, resulting from the material high-term lattice symmetry\cite{ma2023anisotropic}. 

\textbf{Giant Magnetoresistance}
GMR is comprised of three distinct layers. A thin film of non-magnetic conducting metal which is placed between two magnetic layers whose resistance varies depending on the relative orientation of magnetization within the layers. This effect is caused by spin dependent electron scattering; that means when the magnetization between the two layers is aligned only one spin of electrons is able to scatter and thus the overall resistance is minimized. This effect is reversed when the magnetization between the layers is mismatched\cite{khan2021magnetic}.
Spin-dependent transport in vdW GMR junctions with the structure Fe$_3$GeTe$_2$/XTe$_2$/Fe$_3$GeTe$_2$ (X = Pt, Pd) is investigated using first-principles calculations. A GMR of around 2000\% is obtained\cite{li2021current}.

\begin{figure}
\centerline{\includegraphics[width=0.8\textwidth]{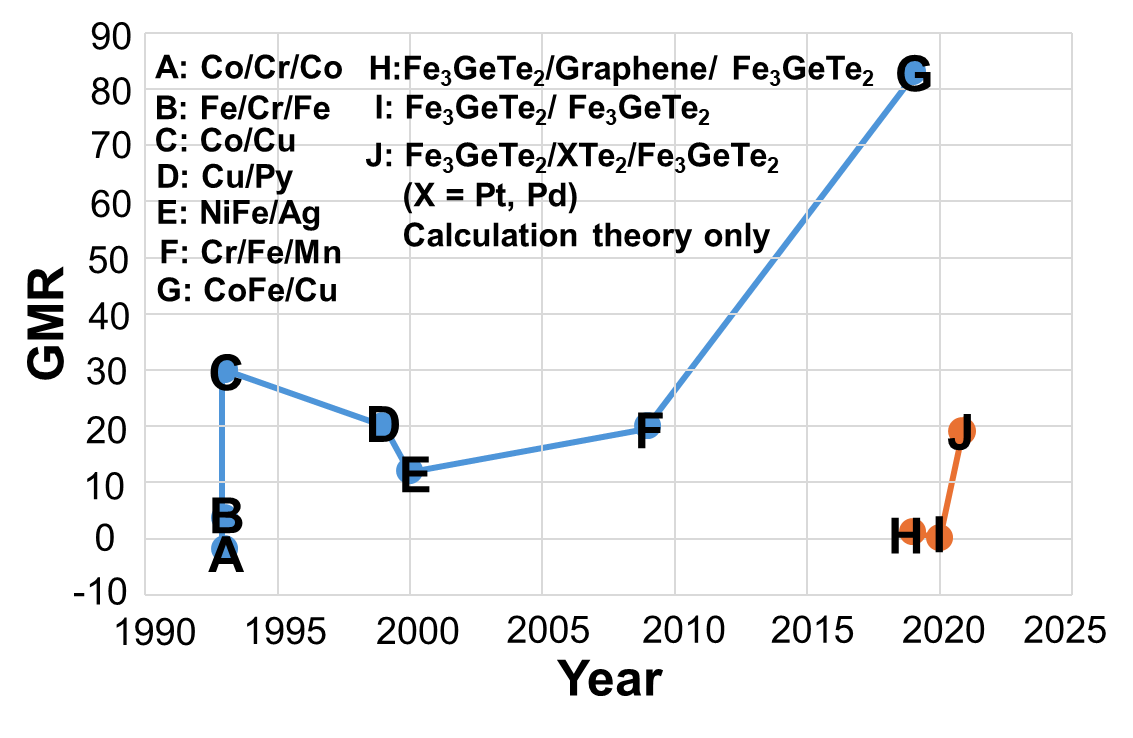}}
\vspace{-20pt}
\caption{GMR ratio of varied systems, including 2D manget system. }
\label{fig-GMR}
\end{figure}

\begin{figure}
\centerline{\includegraphics[width=0.8\textwidth]{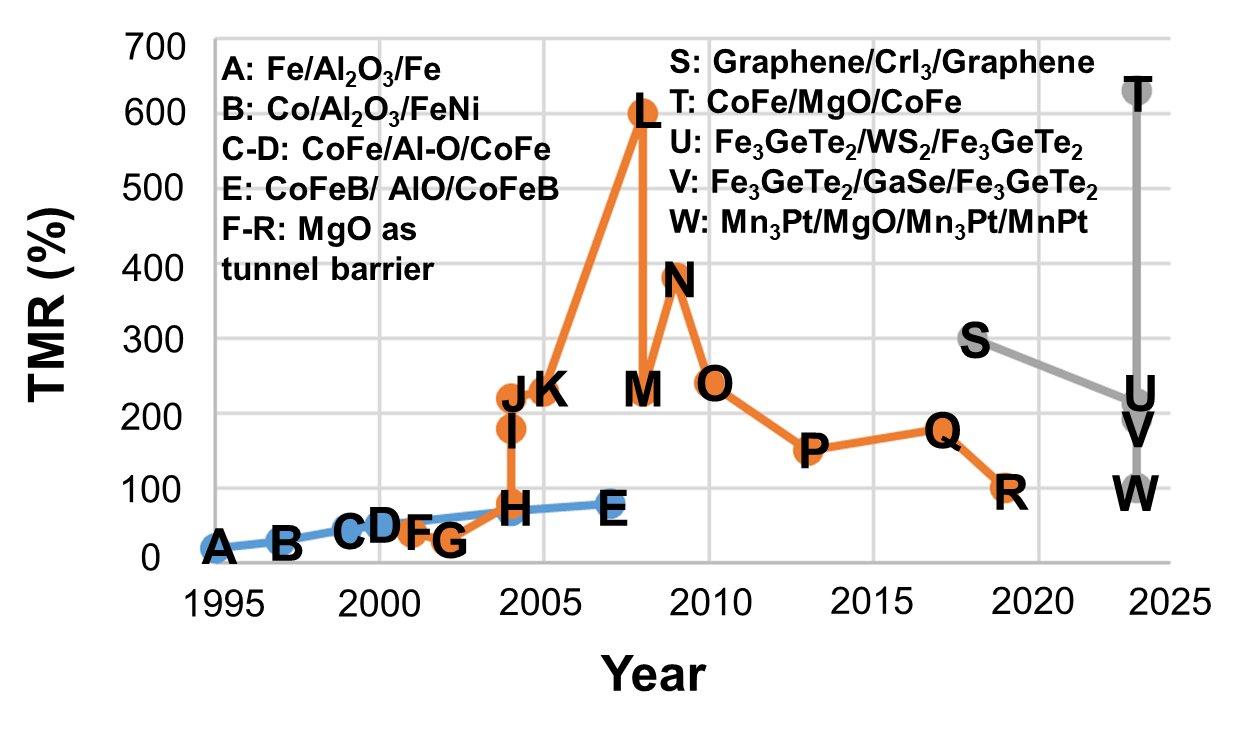}}
\vspace{-20pt}
\caption{Reported TMR ratio since 1995. 2D magnets-based TMR showing relative high ratio.}
\label{fig-TMR}
\end{figure}

\textbf{Tunnel Magnetoresistance}
TMR also referred to as MTJ has a setup similar to GMR except the non-magnetic conducting metal is replaced with an insulator referred to as the tunnel. This tunnel layer acts as a barrier so when the relative orientation of magnetization is aligned between magnetic plates the resistance is low. Conversely, when the relative orientation of magnetization is antiparallel the resistance reaches a peak\cite{khan2021magnetic}. Fig. \ref{fig-TMR} shows reported TMR ratio since 1995. The best on-off ratio is typically limited to 2-6, posing a fundamental challenge in spintronics. The fundamental limit is from (1) selection of right magnetic materials and insulator. (2) interface quality between the materials. Thus, vdW interface would be very promising. Among reported values, vdW heterostructures  relatively high TMR ratio around 2-3. Further progress can be made using ultra-clean vdW interface. 

\subsection{Quantum computing}
%Spin-based quantum computing; Quantum gates using 2D magnets; please add details and figures

Quantum computing, poised to revolutionize computational paradigms, has been a focal point for integrating 2D magnets to realize quantum information processing capabilities. The unique properties of 2D magnets, such as long-range order and tunable magnetic anisotropy, have been harnessed to engineer qubit architectures with enhanced coherence and scalability. One approach is through forming topological superconductivity on 2D magnets, since it was proposed that the dispersing Majorana states could be created at the edges of an island of magnetic adatoms on the surface of an s-wave superconductor \cite{rontynen2015topological,li2016two}. Considering the high-quality interface and strong orbital hybridization \cite{wu2020large} in vdW heterostructures, low-temperature scanning tunneling microscopy and spectroscopy were used to reveal the signatures of one-dimensional Majorana edge modes in vdW superconductor/magnet structure \cite{kezilebieke2020topological}. On the other hand, as exemplified by the recent discoveries of various correlated electronic states in twisted vdW materials, moir\'e patterns can act as a powerful knob to create artificial electronic structures. It was proposed that moir\'e pattern between a vdW superconductor\cite{wu2019induced,han2018investigation} and a monolayer ferromagnet creates a periodic potential modulation that enables the realization of a topological superconducting state that would not be accessible in the absence of the moir\'e\cite{kezilebieke2022moire}. More experiments are needed to carry out to verify these proposals.

\subsection{Quantum sensing and metrology}
%Magnetic field sensing with 2D magnets
%Quantum sensors based on 2D magnetism
%Applications in precision measurement
%Please add examples
In the realm of quantum sensing and metrology, the integration of 2D magnets has ushered in a new era of precision and sensitivity, enabling the measurement of physical quantities such as temperature and time with quantum precision. Harnessing quantum phenomena like spin-dependent transport and quantum entanglement, sensors based on 2D magnets have demonstrated remarkable performance in detecting magnetic fields, gravitational waves, and electromagnetic fields with unprecedented accuracy. As an example, NV centres in diamond have been utilized as quantum sensors owing to the high sensitivity of their spin state to environmental conditions, and defects that are localized near or at the surface of the host material are likely to provide improved sensitivity and accessibility. However, despite progress in the growth of diamond thin films and nanowires, the difficulty of integrating diamond components into devices using traditional lithographic approaches motivates the exploration of defect states in 2D material systems, like h-BN and transition metal dischalcogenides. Moreover, this high-sensitivity quantum sensing holds promise for characterizing nanoscale 2D magnetism. For instance, the NV center in diamond has emerged as a potent probe for static magnetism in 2D vdW materials, offering quantitative imaging at nanoscale spatial resolutions. By employing coherent control of the NV center's spin precession, researchers achieved ultrasensitive, quantitative ac susceptometry of a 2D ferromagnet, CrBr$_3$ \cite{zhang2021ac}.

%\subsubsection{biomedical applications like MRI} Douglas? 

\section{Challenges and opportunities}

\begin{figure}
\centerline{\includegraphics[width=0.9\textwidth]{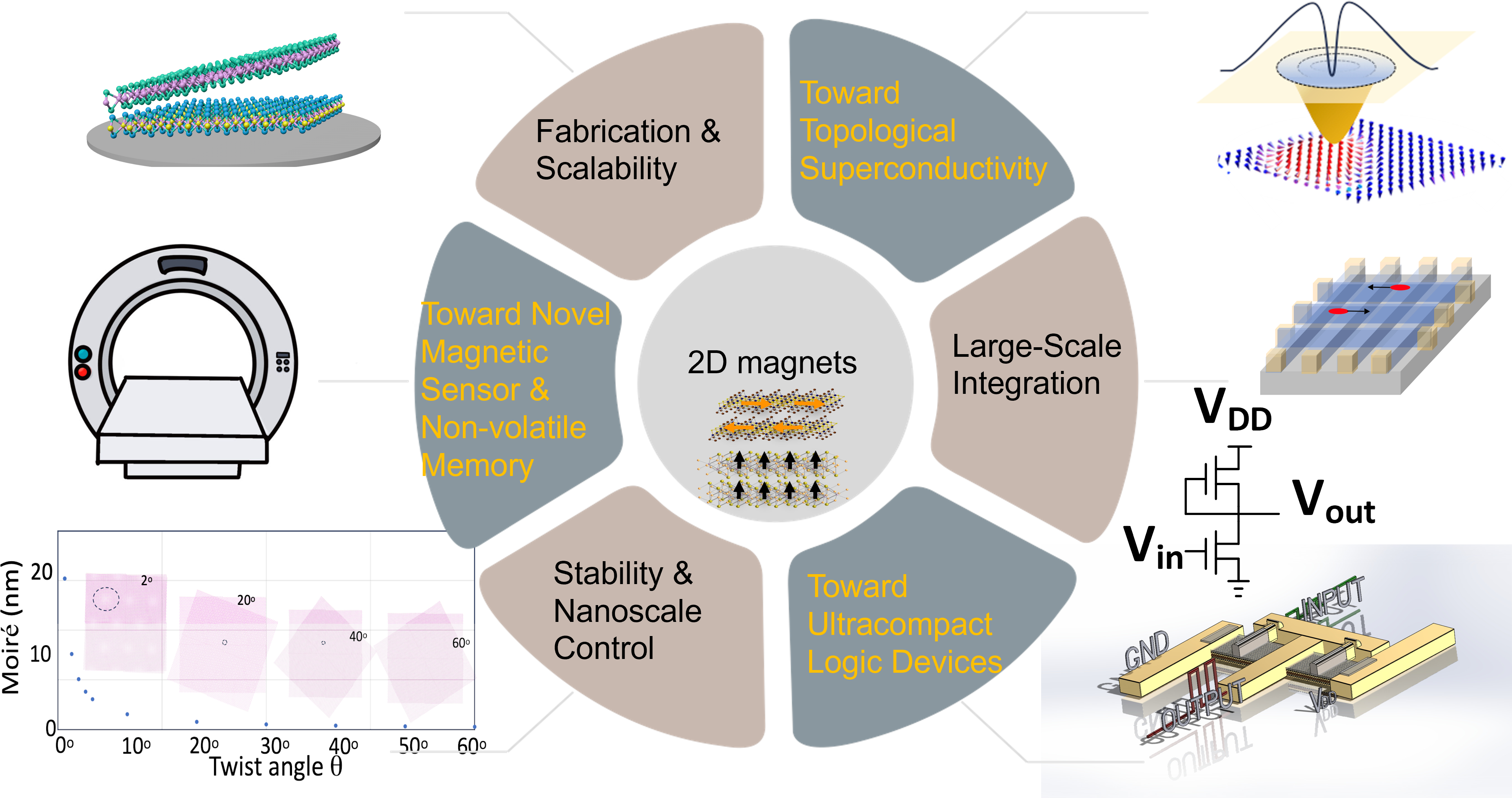}}
\vspace{-20pt}
\caption{Challenges and opportunities in 2D magnets-based quantum devices. Challenges existing in materials and devices level like farication and scalability, stability and nanoscale magnetization conntrol, and large-scale integration. Future applications of these 2D magnets-based quantum devices are topological superconductivity for quantum computing, novel magnetic sensor, non-volatile memory and ultracompact logic devices.}
\label{fig-outlook}
\end{figure}
\subsection{Fabrication and scalability of 2D magnets}
The fabrication and scalability of 2D magnets present significant challenges despite their promising properties. One major hurdle lies in achieving high-quality, large-area 2D materials with uniform magnetic properties (Fig. \ref{fig-outlook}). Current techniques such as conventional mechanical exfoliation using scotch tape and chemical vapor deposition often struggle to consistently produce 2D magnets with controlled thickness and desired magnetic characteristics across a scalable format. Progress has been in MBE grown mm to cm sized Fe$_3$GeTe$_2$ thin films\cite{liu2017wafer,wang2020above}. Moreover, integrating these materials into functional devices requires precise control over layer stacking, interface quality, and maintaining magnetic order at reduced dimensions. Another critical issue is ensuring the stability and durability of 2D magnets under real-world conditions, including environmental factors and operational stresses, which can affect their performance and longevity. Addressing these challenges demands innovative fabrication methods, advanced characterization techniques, and comprehensive understanding of the fundamental physics governing 2D magnetism, ultimately paving the way for practical applications in future technologies.

\subsection{Stability and control of magnetic properties at the nanoscale}
The stability and control of magnetic properties at the nanoscale in 2D magnets represent pivotal challenges and opportunities (Fig. \ref{fig-outlook}) in the field of nanomagnetism. In the case of CrI$_3$, which was extensively studied for its antiferromagnetic coupling and semiconducting properties, its thin films can easily and rapidly disapprear in the air due to the oxidization. Achieving stable magnetic properties requires precise control over factors such as layer thickness, defects, and strain, which can significantly impact magnetic anisotropy and coercivity. Moreover, the ability to manipulate these properties dynamically, for instance, through external stimuli like electric fields or spin currents, opens avenues for developing next-generation magnetic devices with enhanced functionality and energy efficiency. Yet, due to either metallic or insulating properties of most 2D magnets, tuning of carrier density in the system is not easy to achieve. Many more semiconducting magnets are yet to be discovered, which will propel this field forward. Overcoming these challenges involves advancing fabrication techniques to achieve defect-free and reproducible 2D magnet structures, as well as exploring novel approaches to tailor magnetic properties at the nanoscale, thereby unlocking the full potential of 2D magnets in future nanoelectronics and spintronic applications.

\subsection{Integration with existing quantum technologies}
The integration of 2D magnets with existing quantum technologies presents an exciting frontier in materials science and quantum engineering (Fig. \ref{fig-outlook}). By combining the unique magnetic properties of 2D materials\cite{chen2016probing,wu2016negative} with superconductivity or spintronics, researchers aim to create novel devices with enhanced functionalities\cite{zhang20242d}. The coexistence of 2D magnetism and superconductivity in layered materials offers intriguing prospects for exploring new quantum phenomena and potentially uncovering mechanisms that could advance quantum computing and communication technologies. For example, recent progress has been made on topological superconductivity in vdW hetersotructures, where low-temperature scanning tunnelling microscopy and spectroscopy were used to reveal the signatures of one-dimensional Majorana edge modes\cite{kezilebieke2020topological}.  Moreover, 2D topological spin texture like skyrmion can be incorporated to carry the Majorana edge mode\cite{pathak2022majorana}. This interdisciplinary approach promises to unlock new capabilities and applications at the intersection of magnetism and quantum science, paving the way towards more efficient, versatile, and powerful quantum technology. 

\subsection{Opportunities for further research and development}
Opportunities for further research and development in 2D magnets abound (Fig. \ref{fig-outlook}), driven by their unique properties and potential applications across various fields. Exploring new 2D materials and understanding their magnetic behaviors at atomic scales could uncover novel magnetic phenomena and enhance our ability to engineer customized magnetic properties. This can be done integrating artificial intelligence, where the materials properties can be predicted by supervised/unsupervised learning\cite{rhone2023artificial,rhone2020data}. Integrating 2D magnets with other nanoscale materials, such as superconductors or topological insulators, offers exciting prospects for discovering emergent quantum states and advancing quantum information technologies. Additionally, investigating dynamic control mechanisms for manipulating magnetic states in 2D materials, such as SOT or magnetoelectric effects, promises to unlock new functionalities for ultrafast and energy-efficient magnetic devices. For the industry side, 2D magnets offer innovative solutions and advancements across multiple sectors. In electronics, these non-volatile memories based on 2D magnets could significantly enhance the performance and scalability of storage solutions, reducing power consumption and improving data access speeds. Moreover, in sensor technologies, the unique magnetic properties of 2D materials can be leveraged to create highly sensitive and miniaturized magnetic sensors for diverse applications, including biomedical diagnostics, automotive systems, and industrial monitoring. Additionally, the integration of 2D magnets into spintronic devices holds promise for realizing next-generation computing architectures with enhanced processing capabilities and reduced energy consumption. Furthermore, advancements in 2D magnet research could lead to breakthroughs in quantum technologies, enabling the development of more robust and versatile platforms for quantum computing, communication, and sensing. 

\newpage
\textbf{References}
\bibliographystyle{unsrt}
\bibliography{review-2D-magnetic-hetero.bib}

\end{document}